\documentclass[tightenlines,aps,amsmath,showpacs,showkeys,nofootinbib,
superscriptaddress]{revtex4}
\usepackage[dvips]{graphicx}

\begin{document}

\title{On the magic of transforming $\bf{\bar K}$ nuclear broad quasibound 
states into narrow intrinsic decaying states}

\author{A.~Ciepl\'{y}}
\email{cieply@ujf.cas.cz}
\affiliation{Nuclear Physics Institute, 25068 \v{R}e\v{z}, Czech Republic}

\author{A.~Gal}
\email{avragal@vms.huji.ac.il}
\affiliation{Racah Institute of Physics, The Hebrew University, Jerusalem 
91904, Israel} 

\date{\today} 

\begin{abstract}
Comments are made on Akaishi, Myint and Yamazaki's interpretation of 
$\bar K$ quasibound nuclear states as generalized Kapur-Peierls decaying 
states [arXiv:0805.4382]. We argue that these `intrinsic decaying states' 
have little to do with low-energy $\bar K N$ dynamics. 
\end{abstract}

\pacs{13.75.Jz, 21.65.Jk, 21.85.+d} 

\keywords{$\bar K$-nuclear quasibound states}

\maketitle 


\section{Introduction} 
\label{sec:intro}

Akaishi, Myint and Yamazaki (AMY) have argued recently \cite{AMY08}, using 
examples taken from the phenomenology of $\bar K N$ and $\bar K NN$ systems, 
that quasibound states should not be defined by $S$-matrix poles on the 
appropriate Riemann sheet within nonrelativistic coupled-channel potential 
models. Their working example is a $2 \times 2$ coupled channel problem, 
$\bar K N-\pi \Sigma$, in which the potential in the upper $\bar K N$ 
channel is sufficiently strong to generate on its own a bound state that 
becomes quasibound when the two channels are coupled to each other. [This 
is a carricature of the $\Lambda(1405)$ $\bar K N$ quasibound state which 
is identified experimentally by observing a $\pi \Sigma$ resonance shape 
in various reactions.] Increasing the attraction in the upper channel, 
AMY observed that the $\Lambda(1405)$ pole moved to lower energies towards 
the $\pi \Sigma$ threshold, while becoming substantially broader. This large 
width of the quasibound pole state is incompatible, according to AMY, 
with the expected narrowness of the $\bar K N$ spectral shape near the 
$\pi \Sigma$ threshold. Instead, they suggested that Kapur-Peierls 
inspired Intrinsic Decaying States (IDS) replace $S$-matrix pole states 
whenever the width of the latter exceeds some relatively small value. 

In this note we argue that these IDS do not emerge from any proper 
multichannel dynamics for the $\bar K N$ system at low energies, and thus 
IDS are not the correct theoretical construct to use for quasibound states. 
We also show, using a {\it realistic} example from low-energy $\bar K N$ 
phenomenology, that the lower among the two $S$-matrix poles that arise 
naturally in chirally motivated models becomes gradulaly narrow when the 
strength of the $\bar K N$ interaction is beefed up, joining smoothly 
a bound state pole below the $\pi \Sigma$ threshold. Therefore, chirally 
motivated models do not exhibit the pole structure that AMY were bothered by. 

Although the motivation of AMY was apparently to discredit the relatively 
large widths of order 100 MeV found for a $K^-pp$ quasibound state in 
coupled-channel three-body Faddeev calculations by Shevchenko et al. 
\cite{SGM08a,SGM08b}, compared to the smaller width about 60 MeV found in 
the single-channel three-body non-Faddeev calculation by Yamazaki and Akaishi 
\cite{YAk02}, we chose not to enter into argument on this point. 
The AMY paper does not report any new calculation for the $K^-pp$ system 
beyond handwaving in terms of IDS, a concept that is refuted in the present 
note. For this reason we decided not to overdo our criticism of their work.

\section{Kapur-Peierls versus Gamow States}
\label{sec:vs} 

Here we sketch schematically the definitions and properties of Gamow 
states and of Kapur-Peierls states, without specifying the coupled 
channels involved in the physics of the problem. Suffice to state 
that our considerations hold for an effective one-channel Hamiltonian 
$\cal H$ which is energy dependent and is not necessarily hermitian. 

\subsection{Gamow States} 
\label{sec:G} 

Gamow states, corresponding to poles of the $S$ matrix, were introduced 
by Gamow in 1928 to explain decay phenomena such as $\alpha$ decay of 
radioactive nuclei \cite{Gam28}. Gamow states provide a straightforward 
generalization of (normalizable) bound states at real energies to unstable 
(or quasi-) bound states and to resonances at complex energies in terms 
of intrinsic properties of the system and its Hamiltonian $\cal H$. 
Denoting by $\mid{\cal E}_{\rm G}>$ a Gamow state at a complex energy 
${\cal E}_{\rm G}=E_R-{\rm i}\Gamma_R/2$, it satisfies 
${\cal H}\mid{\cal E}_{\rm G}>={\cal E}_{\rm G}\mid{\cal E}_{\rm G}>$, 
with a purely outgoing-wave boundary condition $<r\mid{\cal E}_{\rm G}> 
\propto \exp({\rm i}\sqrt{(2m/{\hbar}^2){\cal E}_{\rm G}}r)$ for 
$r \to \infty$. The time evolution of $\mid{\cal E}_{\rm G}>$ is given by 
\begin{equation} 
\label{eq:tevol} 
\exp(-{\rm i}{\cal H}t/\hbar)\mid{\cal E}_{\rm G}>=\exp(-{\Gamma_R}t/(2\hbar)) 
\times \exp(-{\rm i}E_Rt/\hbar)\mid{\cal E}_{\rm G}>, 
\end{equation} 
providing an exponential decay law with a lifetime $\tau_R=\hbar/\Gamma_R$.  
A unique property of Gamow states is that the transition amplitude from 
a Gamow resonant state at a complex energy ${\cal E}_{\rm G}$ to a scattering 
state of real energy $E>0$ is given by a Breit-Wigner (BW) 
amplitude 
\begin{equation} 
\label{eq:Gamp} 
f_{\rm BW}(E)~\propto~\frac{1}{E-{\cal E}_{\rm G}},  
\end{equation} 
resulting in a BW resonance form of the cross section 
\begin{equation} 
\label{eq:BW} 
\sigma_{\rm BW}(E)~\propto~\frac{1}{(E-E_R)^2+(\Gamma_R/2)^2}. 
\end{equation} 
Eq.~(\ref{eq:BW}) holds only in the immediate neighborhood of the Gamow 
resonance pole, so its validity is limited to {\it narrow} resonances, 
and away from thresholds. To summarize, quoting from a recent reference 
on the properties of Gamow states \cite{Dlm08}: ``Gamow states unify the 
concepts of resonance and decaying particle, and they provide a `particle 
status' for these concepts".

\subsection{Kapur-Peierls States} 
\label{sec:KP} 

Kapur-Peierls (KP) states were introduced in 1938 \cite{KPe38} as eigenstates 
of the Hamiltonian $\cal H$ that are regular at $r=0$ and satisfy a purely 
outgoing-wave boundary condition, with a {\it real} wave number corresponding 
to a {\it given} real incoming energy $E_{\rm KP}$, at a radial distance $r_0$ 
outside the range of the potential. The eigenenergies ${\cal E}_{\rm KP}$ are 
complex, depending parametrically on the real $E_{\rm KP}$. Different choices 
of $E_{\rm KP}$ lead to different sets of eigenenergies 
\{${\cal E}_{\rm KP}$\}. None of these eigenenergies coincide with the pole 
energies of the $S$ matrix and, thus, ${\cal E}_{\rm KP}$ are not related to 
a BW amplitude of the form Eq.~(\ref{eq:Gamp}): 
\begin{equation} 
\label{eq:KPamp} 
f_{\rm BW}(E)~\neq~f_{\rm KP}(E)\propto\frac{1}{E-{\cal E}_{\rm KP}}.   
\end{equation} 
We emphasize that the physical BW amplitude does not depend on the choice of 
$r_0$ and that its complex pole energy ${\cal E}$ does not depend on the 
incoming energy $E_{\rm KP}$. It is worth noting that Peierls' subsequent 
contributions to the subject of resonances hinged exclusively on Gamow states 
\cite{Pei59,GCP76}. In a posthumous publication \cite{DPe97}, Peierls made 
a comment that KP resonances ``are somewhat artificial because they are 
defined with the boundary condition that is correct only at one energy."

\section{Intrinsic Decaying States}
\label{sec:IDS} 

In Eq.~(\ref{eq:KPamp}), $f_{\rm KP}(E)$ is determined by a KP eigenenergy 
${\cal E}_{\rm KP}$ that depends implicitly on the input incoming real 
energy $E_{\rm KP}$. AMY sought to overcome this difficulty by replacing 
the energy argument $E$ by the real energy $E_{\rm KP}$ which serves in 
the outgoing-wave boundary condition to solve for ${\cal E}_{\rm KP}$. 
Schematically, this prescription is expressed as 
\begin{equation} 
\label{eq:fAMY} 
f_{\rm AMY}(E_{\rm KP}) \propto 
\frac{1}{E_{\rm KP}-{\cal E}_{\rm KP}(E_{\rm KP})}. 
\end{equation} 
It is reasonable to assume that the dominant contribution to this amplitude 
arises from the vicinity of the real energy $E_{\rm KP}$ that satisfies 
\begin{equation} 
\label{eq:IDS} 
\Re~{\cal E}_{\rm KP}(E_{\rm KP})~=~E_{\rm KP}. 
\end{equation} 
Eq.~(\ref{eq:IDS}) essentially is how AMY defined the IDS, with a complex 
energy ${\cal E}_{\rm IDS}={\cal E}_{\rm KP}(E_{\rm KP})$ for the solution 
$E_{\rm KP}$ of Eq.~(\ref{eq:IDS}). A more rigorous definition in terms of 
the Hamiltonian ${\cal H}$ is given below. 

\begin{figure} 
\includegraphics[scale=0.7,angle=0]{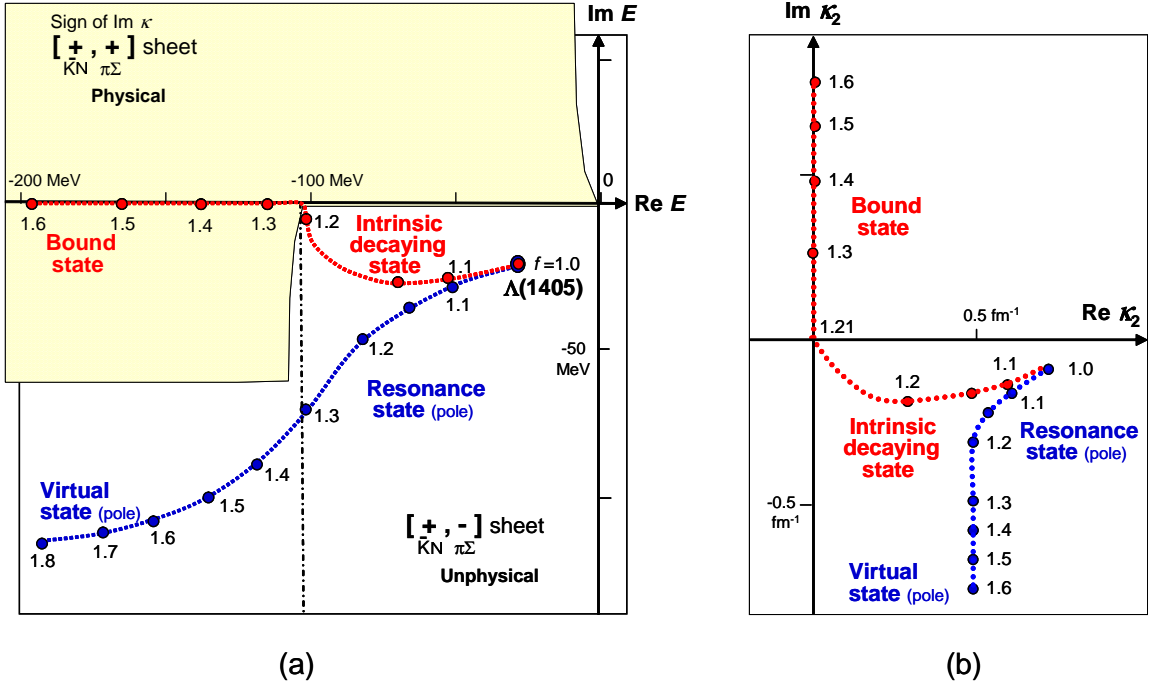} 
\caption{Trajectories of Gamow resonance poles and of IDS: (a) in the complex 
energy plane, and (b) in the $k_{\pi \Sigma}$ (here denoted $k_2$) plane, 
when the strength of the $\bar K N$ interaction is multiplied by a factor $f$ 
marking the points that form the trajectories. Figure taken from 
Ref.~\cite{AMY08}. We thank Professor Yamazaki for permission to reproduce 
the figure.} 
\label{fig:amy08} 
\end{figure} 

The case for IDS is demonstrated in Fig.~\ref{fig:amy08}, taken from 
AMY's paper \cite{AMY08}. Shown on the left-hand side are the trajectories 
of a Gamow resonance pole and of an IDS that approximately coincide at the 
nominal complex energy of the $\Lambda(1405)$, for a standard choice of 
$\bar K N-\pi \Sigma$ coupled-channel Yamaguchi separable interactions. These 
states, for this choice, are located in the fourth quadrant of the complex 
energy plane corresponding to the [$+,-$] sheet, where the signs are those 
of $\Im k_{\bar K N}$ and $\Im k_{\pi \Sigma}$, respectively. The relevant 
portion of this quadrant is bounded from above by the real energy axis from 
the $\pi \Sigma$ threshold up to the $\bar K N$ threshold, about 100 MeV 
higher. The Gamow state appears as a quasibound state in the $\bar K N$ 
channel and as a BW resonance in the $\pi\Sigma$ channel. When the $\bar K N$ 
interaction strength is scaled up by a multiplicative factor $f$ (marked 
along the curves in the figure) from the value $f=1.0$ it assumes for the 
$\Lambda(1405)$, this Gamow pole moves away from the real energy 
axis and its width increases. In contrast, the width of the IDS hardly 
increases upon applying the scaling factor $f$ and ultimately it goes down 
to zero at the $\pi\Sigma$ threshold, joining there smoothly with a bound 
state pole (bound with respect to both thresholds). However, as shown below, 
this is not a consequence of the coupled-channel dynamics. It only reflects, 
as one approaches the $\pi \Sigma$ threshold, the weakening of the imaginary 
part of the effective single-channel $\bar K N$ Hamiltonian used to determine 
${\cal E}_{\rm IDS}$. It accounts for phase space, not for the dynamics. 

In the $\bar K N - \pi \Sigma$ coupled-channel framework discussed by AMY, 
the Schr\"odinger equation is written, using Feshbach's projection operators 
$P$ and $Q$, as
\begin{equation}
PHP~\Psi_P + PVQ~\Psi_Q = E~\Psi_P,~~~~QHQ~\Psi_Q + QVP~\Psi_P = E~\Psi_Q,
\label{eq:cc}
\end{equation}
where $H = T + V$ is the coupled-channel Hamiltonian. Projecting out the 
$Q$ space, one obtains an effective, energy dependent Hamiltonian $H_P$ 
for $\Psi_P$ (identified with the $\bar K N$ wavefunction): 
\begin{equation} 
H_P(E) = PHP + PVQ \frac {1}{E-QHQ+i\epsilon} QVP, 
\label{eq:Hp}
\end{equation}
and similarly an effective, energy dependent $H_Q$ for $\Psi_Q$ 
(identified with the $\pi \Sigma$ wavefunction): 
\begin{equation} 
H_Q(E) = QHQ + QVP \frac {1}{E-PHP+i\epsilon} PVQ. 
\label{eq:Hq} 
\end{equation} 
Note that these effective Hamiltonians are not necessarily hermitian. 
The Gamow resonance states in the [$+,-$] sheet of the complex energy 
plane in Fig.~\ref{fig:amy08} are eigenstates of the coupled channel 
Hamiltonian system Eq.~(\ref{eq:cc}) and are also eigenstates of each 
one of the channel Hamiltonians: 
\begin{equation} 
H_P({\cal E}_{\rm G})\Psi_P={\cal E}_{\rm G}\Psi_P,~~~~ 
H_Q({\cal E}_{\rm G})\Psi_Q={\cal E}_{\rm G}\Psi_Q, 
\label{eq:Hpq} 
\end{equation} 
with a common eigenenergy ${\cal E}_{\rm G}$ and with $\Psi_P$ and $\Psi_Q$ 
satisfying each an outgoing-wave boundary condition. 

The operational definition of IDS, Eq.~(\ref{eq:IDS}), was done in terms of 
a projection onto the $P$ channel only. It is easily shown to be equivalent 
to the following, general definition: 
\begin{equation} 
H_P(\Re {\cal E}_{\rm IDS})\Psi_P={\cal E}_{\rm IDS}\Psi_P. 
\label{eq:HpIDS} 
\end{equation} 
In order to retain a meaning in a coupled-channel formulation, an extension 
onto the $Q$ space is required, which AMY hardly discussed. A sensible 
extension in the spirit of the underlying KP philosophy is to require 
\begin{equation} 
H_Q(\Re {\cal E}_{\rm IDS})\Psi_Q={\cal E}_{\rm IDS}\Psi_Q. 
\label{eq:HqIDS} 
\end{equation}  
However, since $\Re {\cal E}_{\rm IDS}$ is sought between the two 
thresholds, the effective Hamiltonian in Eq.~(\ref{eq:HqIDS}) is {\it real} 
and for the Yamaguchi-type separable potentials used by AMY the resultant 
eigenenergy ${\cal E}_{\rm IDS}$ (if any) is also real, differing from the 
complex value satisfying Eq.~(\ref{eq:HpIDS}). This argument demonstrates that 
the IDS concept cannot be extended satisfactorily from one channel to include 
all the relevant channels in the coupled-channel dynamics. IDS, therefore, 
are not a property of the coupled-channel Hamiltonian, nor of the $S$ matrix 
that determines the physical spectral shapes and cross sections.

\section{Chirally Motivated Models} 
\label{sec:chiral} 

Modern chirally motivated coupled-channel models give rise to {\it two} Gamow 
states that dominate low-energy $\bar K N$ dynamics. For a recent review, see 
Ref.~\cite{HWe08}. One state corresponding to an $I=0$ $\bar K N$ quasibound 
state is generated dynamically from the strongly attractive interaction in 
the $\bar K N$ channel. However, this state cannot be identified with the 
$\Lambda(1405)$. The other Gamow state, on the same Riemann sheet, 
originates from a resonance in the $\pi \Sigma$ channel and it corresponds 
to the physical $\Lambda(1405)$. Here we would like to follow the movement 
of these two poles in the complex energy plane upon changing the strength 
of the interaction in a chirally motivated coupled-channel model developed 
recently by one of us \cite{CSm07}. 

\begin{figure} 
\includegraphics[scale=0.7,angle=0]{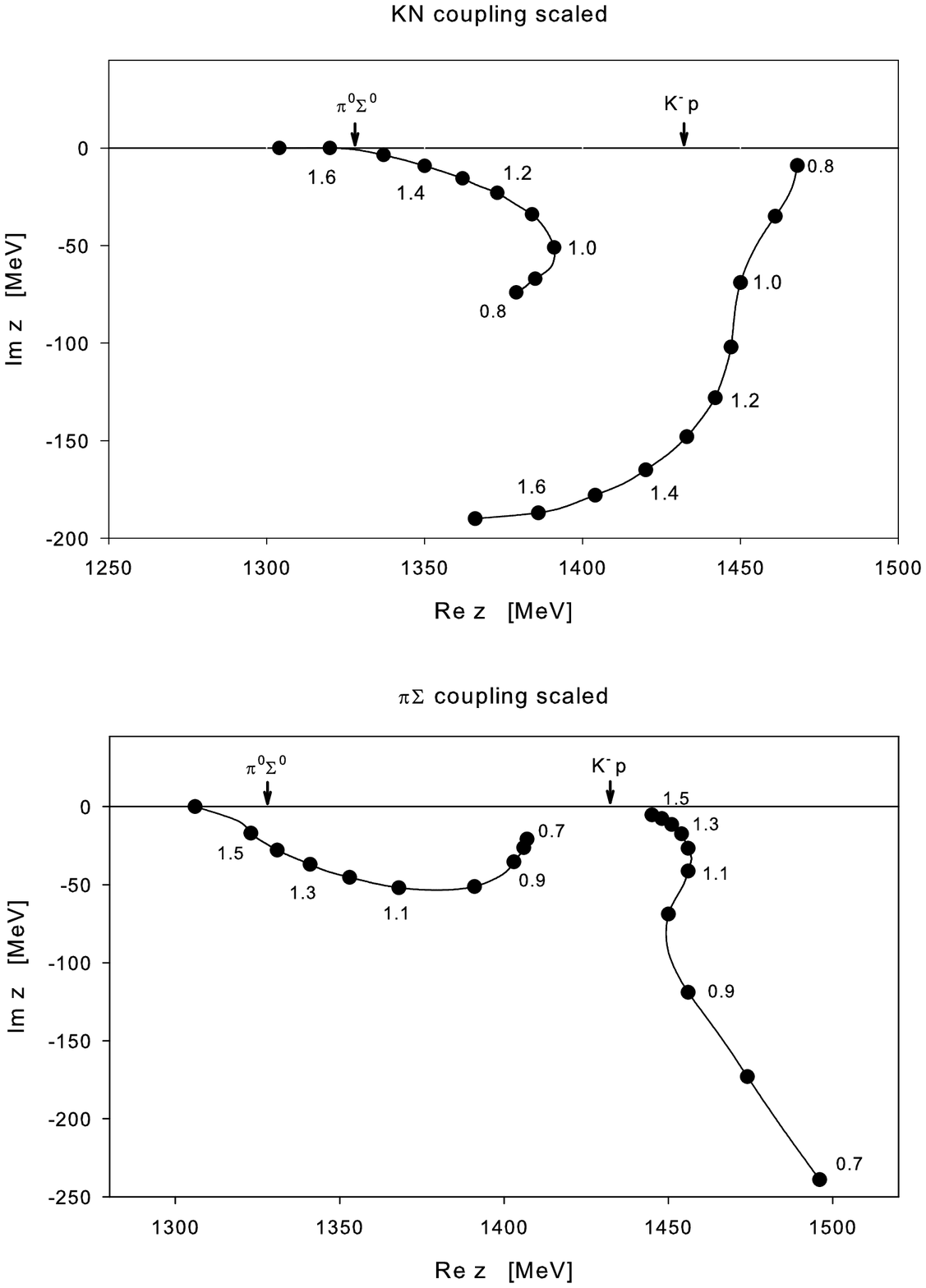} 
\caption{Trajectories of Gamow poles in the complex energy (z) plane on the 
Riemann sheet [$\Im k_{\bar K N},\Im k_{\pi \Sigma}$] = [$+,-$] using the 
10-channel model of Ref.~\cite{CSm07}. The upper (lower) part depicts the 
motion upon scaling the $\bar K N$ ($\pi\Sigma$) interaction strengths by 
a multiplicative factor marking the points that form the trajectories. 
The $\pi^0\Sigma^0$ and $K^-p$ thresholds are marked by arrows.} 
\label{fig:poles} 
\end{figure} 

The model consists of 10 coupled channels, made out of the two-body systems 
$\bar K N, \pi\Sigma, \pi\Lambda, \eta\Lambda, \eta\Sigma, K\Xi$ with zero 
total charge. It fits well all the low-energy $K^-p$ scattering and reaction 
data except for the perennially irreproducible $1s$ atomic width, and it 
reproduces reasonably well the $\pi\Sigma$ spectrum shape which is identified 
with the $\Lambda(1405)$ resonance. To be specific, we use the parameter set 
that gives $\sigma_{\pi N} = 40$ MeV (see Table 2 in Ref.~\cite{CSm07}). 
Other parameter sets that fit the data equally well produce similar trends 
to that discussed below. The model yields two Gamow poles in the fourth 
quadrant of the [$+,-$] Riemann sheet, with composition dominated by 
$\pi\Sigma$ and $\bar K N$ channels. These pole positions, at 
$(E_R,-{\rm i}\Gamma_R/2)=(1391,-{\rm i}51),~(1450,-{\rm i}69)$ (in MeV), 
respectively, are shown in Fig.~\ref{fig:poles} together with the trajectories 
followed by these poles upon multiplying the $\bar K N$ ($\pi\Sigma$) 
interactions by a scaling factor as indicated in the upper (lower) part. 
The $\Lambda(1405)$ resonance corresponds to the lower pole at 
$(1391,-{\rm i}51)$ MeV. The upper pole appears in this model above the 
$\bar K N$ threshold, as it does in other models (see Fig. 8 in 
Ref.~\cite{HWe08} for a compilation of results from various chiral models), 
and it is more likely to be associated with $\bar K$ quasibound states 
in nuclei. Fig.~\ref{fig:poles} shows that this upper pole reacts 
differently to scaling of various pieces of the interactions: 
increasing the strength of the $\bar K N$ interactions, it drifts to 
lower energies below the $\bar K N$ threshold while becoming very broad, 
whereas increasing the strength of the $\pi\Sigma$ interactions, it remains 
above the $\bar K N$ threshold and its width becomes vanishingly small. 
In contrast, the lower pole drifts to lower energies, approaching the real 
axis, and for a sufficiently strong coupling it forms a bound state below 
the $\pi\Sigma$ threshold. A similar dependence of pole positions on scaling 
factors was observed in a recent study of Gamow states within a simplified 
model used to test phenomenological methods of extracting resonance 
parameters in meson-nucleon reactions \cite{SSL08}. While this study presents 
the behavior of resonance poles in a more general context, here we employed 
a realistic multichannel model that is based on chiral dynamics. We conclude 
that Gamow states associated with low-energy $\bar K N$ phenomenology display 
a more subtle pattern than the classification made by AMY.

\section{Concluding Remarks} 
\label{sec:concl} 

In this brief note we argued against the applicability of the IDS 
concept introduced by AMY to describe low-energy $\bar K N$ dynamics. 
Gamow resonance states and Gamow quasibound states, associated with Gamow 
poles, are the only quantum states that provide a proper generalization 
of normalizable bound states. Gamow states are nonrenormalizable eigenstates 
of the multichannel Hamiltonian ${\cal H}$ and satisfy eigenstate 
equations with outgoing-wave boundary conditions, Eqs.~(\ref{eq:Hpq}), 
in each channel. Gamow states are independent of any reaction mechanism 
by which one seeks to establish such resonances or quasibound states. To 
fit and interpret production or formation reaction cross sections in terms 
of quantum states that are intrinsic property of ${\cal H}$, obviously
one needs to superimpose the constraints of phase space which are specific 
to that given reaction. It is wrong, however, to incorporate phase space 
constraints imposed by the reaction which generates such quantum states 
into their definition. 

IDS are {\it not} eigenstates of ${\cal H}$ in {\it all} the relevant 
channels. AMY conjecture that channel $P$ provides the doorway for 
forming a dynamical entity in the reaction they chose to analyze, 
and that's why they geared the IDS to satisfy an eigenstate equation, 
Eq.~(\ref{eq:HpIDS}), in channel $P$. Suppose that we conjecture that 
channel $Q$ provides the doorway for forming the same dynamical entity 
in a different reaction; are we then justified to define IDS by 
satisfying an eigenstate equation in channel $Q$? If we do, the two IDS 
will be different from each other and would not qualify to describe the 
coupled channel dynamics. The example in Sect.~\ref{sec:chiral} of two 
dynamical poles, one arising from $P$-channel interactions while the other 
one from $Q$-channel interactions, provides some justification to exploring 
more than one production/formation mechanism in the study of $\bar K$ 
nuclear quasibound states. For these quasibound states to reflect the 
coupled-channel Hamiltonian dynamics, they must arise with the same 
eigenenergy in all channel spaces, something that IDS are unable to deliver 
in few-body systems where coupled-channel dynamics plays a crucial role.

\begin{acknowledgments} 
This work was supported in part by the GA AVCR grant IAA100480617 and by 
the Israel Science Foundation grant 757/05. 
\end{acknowledgments}

\end{document}